\documentclass{amsart}
\usepackage{amsmath,amsfonts}
\usepackage{a4}

\newtheorem{lemma}{Lemma}

\newtheorem{theorem}{Theorem}

\newcommand{\bp}{\mathbf{p}}

\newcommand{\bx}{{x}}
\newcommand{\by}{{y}}

\newcommand{\gH}{\mathfrak{H}}
\newcommand{\gS}{\mathfrak{S}}
\newcommand{\gSt}{\mathfrak{S}_1^{P^0}}
\renewcommand{\S}{\mathcal{S}}
\newcommand{\E}{\mathcal{E}}
\newcommand{\F}{\mathcal{F}}
\newcommand{\Er}{\mathcal{E}_{{\mathrm{red}}}}

\renewcommand{\phi}{\varphi}

\newcommand{\cz}{\mathbb{C}} 
\newcommand{\rz}{\mathbb{R}} 
\newcommand{\R}{\mathbb{R}} 

\newcommand{\Hh}{\mathbb{H}}

\newcommand{\dK}{{\int_{-\infty}^{\infty}d\eta}}
\newcommand{\ide}{\frac 1{D^0 + i \eta}}

\newcommand{\vp}{\varphi}
\newcommand{\alr}{{\alpha_{\mathrm{r}}}}
\newcommand{\rr}{{\rho_{\mathrm{r}}}}

\newcommand{\supp}{{\mathrm{supp}}}

\newcommand{\hr}{{\hat \rho}}

\newcommand{\alp}{\boldsymbol{\alpha}}
\newcommand{\wto}{\rightharpoonup}

\def\tr{\mathop{\rm tr}\nolimits} 
\def\Tr{{\rm tr}_{\cz^4}}
\def\str{\mathop{\rm tr}\nolimits} 

\title[Self-consistent polarized vacuum]{Self-consistent solution for the polarized vacuum
in a no-photon QED model}

\thanks{The authors are thankful to Robert Seiringer and Vladimir M. Shabaev
for valuable comments.
    They
     acknowledge support through the European Union's IHP
    network Analysis \& Quantum HPRN-CT-2002-00277. E.S. acknowledges 
    support from the Institut Universitaire de France}

\author[C. Hainzl]{Christian Hainzl}
\address{CEREMADE, Universit\'e
  Paris-Dauphine, Place du Mar\'echal de Lattre de Tassigny, F-75775 Paris
  Cedex 16, France \& Laboratoire de Math\'ematiques
Paris-Sud-Bat 425, F-91405 Orsay Cedex} \email{hainzl@ceremade.dauphine.fr}

\author[M. Lewin]{Mathieu Lewin}
 \address{CEREMADE, Universit\'e
  Paris-Dauphine, Place du Mar\'echal de Lattre de Tassigny, F-75775 Paris
  Cedex 16, France.}
  \email{lewin@ceremade.dauphine.fr}

\author[E. S\'er\'e]{Eric S\'er\'e}
 \address{CEREMADE, Universit\'e
  Paris-Dauphine, Place du Mar\'echal de Lattre de Tassigny, F-75775 Paris
  Cedex 16, France.}
  \email{sere@ceremade.dauphine.fr}

\begin{document}


\begin{abstract}
We study the Bogoliubov-Dirac-Fock model introduced by Chaix and Iracane ({\it J. Phys. B.}, 22, 3791--3814, 1989) which is a mean-field theory deduced from no-photon QED. The associated functional is bounded from below. In the presence of an external field, a minimizer, if it exists, is interpreted as the polarized vacuum and it solves a self-consistent equation. 

In a recent paper, we proved the convergence of the iterative fixed-point scheme naturally associated with this equation to a global minimizer of the BDF functional, under some restrictive conditions on the external potential, the ultraviolet cut-off $\Lambda$ and the bare fine structure constant $\alpha$. In the present work, we improve this result by showing the existence of the minimizer by a variational method, for any cut-off $\Lambda$ and without any constraint on the external field. 

We also study the behaviour of the minimizer as $\Lambda$ goes to infinity and show that the theory is ``nullified" in that limit, as predicted first by Landau: the vacuum totally cancels the external potential. Therefore the limit case of an infinite cut-off makes no sense both from a physical and mathematical point of view.

Finally, we perform a charge and density renormalization scheme
applying simultaneously to all orders of the fine structure constant $\alpha$, on a simplified model where the exchange term is neglected. 
\end{abstract}

\maketitle

\section{Introduction\label{s1}}
Despite the incredible predictive power of Quantum Electrodynamics (QED) its
description in terms of perturbation theory restricts its range of applicability.
In fact
a mathematical consistent formulation is still unknown. We want to make a tiny step
in that direction. 

Following ideas of Chaix and Iracane \cite{Chaix}, we study in this paper a model for the polarized vacuum in a Hartree-Fock type approximation. This so-called Bogoliubov-Dirac-Fock (BDF) model has been derived from no-photon QED in \cite{Chaix} as a possible cure to the fundamental problems associated with standard relativistic quantum chemistry calculations. 

The vacuum polarization (VP) is, quoting \cite{FE}, ``one of the most interesting of the phenomena predicted by contemporary quantum electrodynamics". Although it plays a minor role in the calculation of the Lamb-shift for the ordinary hydrogen atom (comparing to other electrodynamic phenomena), it is important for High-$Z$ atoms \cite{MPS,Sh} and even plays a crucial role for muonic atoms \cite{FE,GRS}. It also explains the production of electron-positron pairs, observed experimentally in heavy ions collision \cite{ABHTS,RMG,KS2,Seipp,FS}.

In \cite{Chaix}, Chaix and Iracane noticed that the vacuum polarization effects are ``necessary for the internal consistency of the relativistic mean-field theory and should therefore be taken into account in proper self-consistent calculations, independently of the magnitude of the physical effects" \cite[page 3813]{Chaix}. Taking into account these effects, they restricted the no-photon QED Hamiltonian (normal-ordered with respect to the free electrons and positrons) to Bogoliubov transformations of the free vacuum. This allowed them to obtain a bounded-below energy, a property which is a huge advantage compared to the usual Dirac-Fock theory \cite{DF}: the Dirac-Fock energy is unbounded from below, which is the cause of important computational \cite{Chaix,Chaix_th} and theoretical \cite{ES,Pat,ES3,ES4} problems .

In this paper, we show the existence of a global minimizer for the Bogoliubov-Dirac-Fock functional of Chaix-Iracane in the presence of an external field, which is interpreted as the polarized vacuum. This vacuum is represented by a projector of infinite rank which solves a \emph{self-consistent equation}: it is the projector on the negative eigenspace of an effective mean-field Dirac operator taking into account the vacuum polarization potentials. This equation naturally leads to an iterative fixed-point procedure for solving it. In a previous work \cite{HLS1}, we proved the convergence of such an iterative scheme to a global minimizer of the BDF functional, but under some assumptions on the external field and the ultraviolet cut-off. Our goal here is to show the existence of a minimizer without any restriction, by means of  a direct -- non constructive -- minimization argument.

In the case where no external field is present, the free vacuum is
already known to be a minimizer of the BDF energy, as shown by
Chaix-Iracane-Lions \cite{CIL} and Bach-Barbaroux-Helffer-Siedentop
\cite{BBHS}. In \cite{BBHS}, an external field is also considered but
vacuum polarization is neglected: the model studied there is thus very different from the one considered by Chaix-Iracane in \cite{Chaix} and in the present paper.

Of course the vacuum case is only a first step in the study of the Chaix-Iracane model. In order to consider atoms and molecules, one has to minimize the BDF energy in a fixed charge sector, a much more complicated problem from a mathematical point of view. A minimizer would then solve a self-consistent equation which takes the form of the usual \emph{unprojected Dirac-Fock equations}, perturbed by the vacuum polarization potentials.

To deal with divergencies, we impose a ultraviolet momentum cutoff $\Lambda$. Our only restriction on $\Lambda$
is its finiteness. Additionally we study the behaviour of our solution when $\Lambda\to\infty$ and show that the model becomes meaningless since the vacuum density totally cancels the external potential. In physics, this ``nullification" of the theory as the cut-off diverges has been first predicted by Landau {\it et al.} \cite{Lan86,Lan84,Lan89,Lan100} and later thoroughly studied by Pomeranchuk {\it et al.} \cite{Pom}.

We also discuss a simplified model in more detail,
neglecting the exchange energy. For the corresponding
self-consistent solution we perform a fully -- to any order in the coupling constant $\alpha$ -- consistent charge renormalization scheme.
This procedure has already been performed in perturbation theory by means of Feynman diagrams, see e.g. \cite[page 194]{RGA} and \cite{Hamm}.
In particular, we recover the well-known fact \cite{Lan84,Lan86,Lan89} (see also, e.g., \cite[Eq. $(7.18)$]{IZ}) that the physical (renormalized) coupling constant $\alpha_{\rm r}$  is related to the bare $\alpha$ by a relation of the form
\begin{equation}
\alr = \frac \alpha{1+\alpha B_\Lambda}.
\label{ren_alpha0}
\end{equation} 
where $B_\Lambda\sim_{\Lambda\to\infty}2/(3\pi)\log\Lambda$. Therefore the limit case of an infinite cut-off appears as unphysical \cite{Lan86} since it would correspond to $\alpha_{\rm r}=0$, which means no more electrostatic interactions.

The paper is organized as follows. In the next section, we recall the BDF model. Our main existence result is stated in Section 3, together with the behaviour of the solution as $\Lambda\to\infty$. In Section 4, we study the reduced model and interpret the self-consistent equation thanks to a renormalization of the charge and the density. Finally, the last section is devoted to the proof of our main results.

\section{The Bogoliubov-Dirac-Fock model}
For the sake of clarity, we first briefly recall the Bogoliubov-Dirac-Fock (BDF) model introduced by Chaix-Iracane in \cite{Chaix} and studied in \cite{HLS1}. Details can be found in \cite{HLS1}.

We use relativistic units $\hbar = c = 1$, set the particle mass equal to one and
$\alpha = e^2/(4\pi)$. We emphasize that
in the first part $e$ represents the {\em bare} charge of the electron.
We assume the presence of an external field
$\varphi = n\ast \frac{1}{|\cdot|}$ describing one or more extended nuclei with
overall charge density $n(x)$. We do not assume in this work that $n$ is a non-negative function, since our model allows to treat the vacuum interacting with both matter and antimatter.
We denote by $D^0 = \alp \cdot \bp + \beta$ the free Dirac operator and by 
$D^\vp := D^0 - \alpha \vp$ the Dirac operator with external potential.
Throughout the paper we use the notation $\chi_{{(-\infty,0)}}(H)$ to denote
the projector on the negative spectral subspace of $H$. In the physical literature 
$\chi_{{(-\infty,0)}}(H)$ is often denoted as $\Lambda^-(H)$. 

When the external field is 
not too strong, a good approximation is to use the Furry picture \cite{Fu} in 
the Lamb-shift calculations of atomic bound states (see, e.g., \cite{MPS,Sh}). This means that, in order to evaluate corrections due to Vacuum Polarization, the dressed vacuum is represented by the projector associated with the negative spectrum of the Dirac operator with external potential $D^\vp$
$$P^\varphi=\chi_{(-\infty;0)}(D^\varphi).$$
In reality, the 
polarized vacuum modifies the electrostatic field, and the virtual 
electrons react to the corrected field. This remark naturally leads to a 
self-consistent equation for the dressed vacuum of the form
$$P_{\rm scf}=\chi_{(-\infty;0)}(D^\varphi+V_{{\rm eff}})$$
where $V_{\rm eff}$ is an effective potential already including the Vacuum Polarization potentials.
The BDF model \cite{Chaix} allows to derive such an effective potential $V_{{\rm eff}}$ in a {\em self-consistent} way, $P_{\rm scf}$ being interpreted as a minimizer in the class of Bogoliubov transformations of the free vacuum $P^0=\chi_{(-\infty;0)}(D^0)$.

In practice, $V_{\rm eff}$ can be computed by a fixed point iterative procedure studied in details in \cite{HLS1}.
If one starts the procedure from 
$P^0$, the first iteration gives $P^{\varphi}$, and this explains why 
the Furry picture is a good approximation. But 
corrections to the Furry picture are necessary for high accuracy 
computations of electronic levels near heavy nuclei. These 
corrections can be interpreted as the second iteration in a Banach 
fixed-point algorithm (see, e.g., \cite[section 8.2]{MPS}).

Self-consistent equations leading to a fixed-point iterative scheme are well-known and widely used in full QED. The solutions of the Schwinger-Dyson equations \cite{Sch4,Dy} involving the different four-dimensional Feynman propagators are usually found by means of perturbation theory. Our approach for the special case of the Hartree-Fock theory without photon studied in this paper is mathematically rigorous, non-perturbative and works for any charge $Z$ of
the external potential. 

\bigskip

The momentum cutoff $\Lambda$ is implemented in the
Hilbert space 
\begin{equation}
\gH_\Lambda = \{ f \in L^2(\rz^3, \cz^4)\ |\ \supp \hat f \subset B(0,\Lambda)\},
\label{cutoff}
\end{equation} 
that is the space of spin valued functions whose Fourier transform
has support inside a ball with radius $\Lambda$. Such a sharp cut-off does not allow to keep gauge invariance when photons are present. Since we neglect photons, we shall however use \eqref{cutoff} for simplicity. 

The space $\gH_\Lambda$ can be decomposed as a direct sum of the negative and positive subspaces of the free Dirac operator $D^0$, i.e. $\gH_\Lambda=\gH_-^0\oplus\gH_+^0$ where $\gH_-^0=P^0\gH_\Lambda$ and $\gH_+^0=(1-P^0)\gH_\Lambda$, $P^0=\chi_{(-\infty;0)}(D^0)$. The Fock space $\F$ is built upon this splitting as usual \cite{Chaix,Thaller}:
$$\F:=\bigoplus_{n,m=1}^\infty \F^{(n)}_+\otimes\F^{(m)}_-,$$
where $\F^{(n)}_+:=\bigwedge_{i=1}^n\gH_+^0$ is the $n$-electron state subspace,  $\F^{(m)}_-:=\bigwedge_{j=1}^mC\gH_-^0$ is the $m$-positron state subspace, and $\F^{(0)}_+=\F^{(0)}_-=\cz$. Here $C$ is the charge conjugation operator \cite{Thaller}.
The bare annihilation operators for electrons $a_0(f)$ and positrons $b_0(f)$ are then defined in the usual way \cite{Chaix,Thaller}, for any $f\in\gH_\Lambda$. The field operator reads
$$\Psi(f)=a_0(f)+b_0^*(f).$$
The free vacuum $\Omega_0=1\in\cz\subset\F$ is caracterized up to a phase by the properties $a_0(f)\Omega_0=b_0(f)\Omega_0=0$ for any $f\in\gH_\Lambda$, and $\|\Omega_0\|_{\F}=1$.

Let us now define the BDF class in the Fock space. Given a new (dressed) projector $P$, we define the dressed annihilation operators by $a_P(f)=\Psi((1-P)f)$ and $b_P(f)=\Psi^*(Pf)$. The associated dressed $\Omega_P$ is a state in the Fock space such that $a_P(f)\Omega_P=b_P(f)\Omega_P=0$ for any $f\in\gH_\Lambda$, and $\|\Omega_P\|_\F=1$.
By the Shale-Stinespring Theorem \cite{ShSt}, such an $\Omega_P$ is known to exist and is unique up to a phase, if and only if $P-P^0\in \gS_2(\gH_\Lambda)$, the space of Hilbert-Schmidt operators on $\gH_\Lambda$ (see also \cite{KS}). The state $\Omega_P$ can be expressed as a rotation of the free vacuum, $\Omega_P=\mathbb{U}\Omega_0$, $\mathbb U$ being called a Bogoliubov transformation. An explicit formula for $\Omega_P$ can be found in lots of papers \cite{Thaller,KS,Rui,SS,Seipp,FS}. The BDF class is therefore the subset of $\F$
$$\mathcal{B}:=\left\{\Omega_P\ |\ P \text{ orth. projector},\ P-P^0\in\gS_2(\gH_\Lambda)\right\}.$$

The charge of $\Omega_P$ can be easily computed
\begin{eqnarray}
\label{charge}
\langle\Omega_P|\mathcal{Q}|\Omega_P\rangle & = & \tr(P^0(P-P^0)P^0)+\tr((1-P^0)(P-P^0)(1-P^0))\\
 & = & \tr(Q^{--})+\tr(Q^{++})\nonumber
\end{eqnarray} 
where $Q=P-P^0\in\gS_2(\gH_\Lambda)$ and $Q^{--}=P^0QP^0$, $Q^{++}=(1-P^0)Q(1-P^0)$. In \eqref{charge}, 
$\mathcal{Q}$ is the usual charge operator acting on the Fock space $\F$ \cite[Eq. $(10.52)$]{Thaller},
$$\mathcal{Q}=\sum_{i\geq1} a_0^*(f_i^+)a_0(f_i^+) -\sum_{i\geq1}b_0^*(f_i^-)b_0(f_i^-),$$
$(f_i^+)_{i\geq1}$ and $(f_i^-)_{i\geq1}$ being respectively orthonormal basis of $\gH_+^0$ and $\gH_-^0$.

Due to \eqref{charge}, we have introduced in \cite{HLS1} the notion of
$P^0$-trace class operators.
We say $A\in \gS_2(\gH_\Lambda)$ is $P^0$-trace class
if the operators $A^{++}:=(1-P^0)A(1-P^0)$ and $A^{--}:=P^0 A P^0$ are
trace-class ($\in \gS_1(\gH_\Lambda)$),
and we define the $P^0$-trace of $A$ by
\begin{equation}
\tr_{P^0} A = \tr A^{++} + \tr A^{--}.
\end{equation}
Notice, if $A$ is even trace-class
then  $\tr_{P^0} A = \tr A$. In the following, we denote by $\gSt(\gH_\Lambda)$ the set of all $P^0$-trace class operators. Remark that by definition $\gSt(\gH_\Lambda)\subset\gS_2(\gH_\Lambda)$.

We have shown in \cite[Lemma 2]{HLS1} that any difference of two projectors satisfying the Shale-Stinespring criterion, $Q=P-P^0 \in \gS_2(\gH_\Lambda)$, 
is automatically
in $\gSt(\gH_\Lambda)$. The charge $\langle\Omega_P|\mathcal{Q}|\Omega_P\rangle =\tr_{P^0}(Q)$ is therefore a well-defined number which indeed is always an integer, as proved in \cite[Lemma 2]{HLS1}.
The $P^0$-trace is an adequate tool for describing charge sectors, without using the explicit expression of $\Omega_P$ which can be found in the literature.

In this paper, we study the case of the vacuum: namely we want to show the existence of a BDF state $\Omega_P\in\mathcal{B}$ with \emph{lowest energy}, which we call a \emph{BDF-stable vacuum}. For a small external field, this vacuum will not be charged but if the external field is strong enough, we could end up with a charged vacuum, $\langle\Omega_P|\mathcal{Q}|\Omega_P\rangle=\tr_{P^0}(P-P^0)\neq0$. In order to study atoms or molecules, one has to minimize the energy in different charge sectors 
$$\mathcal{B}_N:=\left\{\Omega_P\in\mathcal{B}\ |\ \langle\Omega_P|\mathcal{Q}|\Omega_P\rangle=N\right\}\subset\mathcal{B}.$$
In this case, as explained in \cite[section 4.2]{Chaix} (see also \cite[Remark 6]{HLS1}), the electronic orbitals will solve the unprojected Dirac-Fock equations, perturbed by the vacuum polarization potentials. It is our goal to study this constrained minimization problem in the near future.

\bigskip

According to Chaix and Iracane \cite[Formula $(4.1)$]{Chaix}, the energy of a state $\Omega_P$ is defined using the renormalized Hamiltonian, acting on the Fock space $\F$,
\begin{equation}
  \label{rh}
  \Hh = \int dx\,: \Psi^*(x)
  D^{\vp} \Psi(x) :_{P^0}
  + \frac{\alpha}2 \int dx \int dy
  \frac{:\Psi^*(x)\Psi(x)\Psi^*(y)\Psi(y):_{P^0}}{|\bx - \by|}
\end{equation}
where $\Psi(x)=\sum_{i\geq1}\Psi(f_i)f_i(x)$,  $(f_i)_{i\geq1}$ being an orthonormal basis of $\gH_\Lambda$. The choice of the normal ordering with respect to $P^0$ corresponds to subtracting the energy of the free Dirac sea $P^0$
and the interaction potentials involving $P^0$. We emphasize that by this choice we make the assumption that the free vacuum is unobservable, as done by Dirac \cite{D1,D2}, Heisenberg \cite{Hei} and Weisskopf \cite{Weis} (see also \cite{HS}). In principle, other choices could be made \cite{LS}.

Evaluating the expectation value of $\Omega_P$, we obtain \cite[Appendix]{HLS1}
\begin{equation}
\langle \Omega_P|\Hh |\Omega_P\rangle=\mathcal{E}(Q)
\label{expectation}
\end{equation} 
where $Q=P-P^0\in\gS_1^{P^0}(\gH_\Lambda)$ and $\E$ is the Bogoliubov-Dirac-Fock energy
\begin{multline}
\label{energy}
\mathcal{E}(Q)=\str_{P^0}(D^0 Q)-\alpha D(\rho_Q,n) +\frac{\alpha}{2}D(\rho_Q, \rho_Q)
-\frac{\alpha}{2}\int\!\!\!\int
\frac{|Q(x,y)|^2}{|\bx-\by|}dx\,dy.
\end{multline}
Here $\rho_Q(\bx) = \Tr Q(\bx,\bx)$ and
$$D(f, g) = 4\pi\int_{\R^3}\frac{\overline{\widehat{f}(k)}\widehat{g}(k)}{|k|^2}dk.$$
Notice that the density $\rho_Q$ is well defined due to the ultraviolet cut-off \cite[Eq. (9)]{HLS1}, and that $D(f,g)=\iint_{\R^6}\frac{f(x)g(y)}{|x-y|}dx\,dy$ when $f$ and $g$ are smooth enough.

As this is seen from \eqref{expectation}, the energy of $\Omega_P$ only depends on $Q=P-P^0$, which is interpreted as the \emph{renormalized one-body density matrix} of $\Omega_P$.

\section{Existence of a BDF-stable polarized vacuum}
Following a usual method for Hartree-Fock type theories \cite{Lieb,Bach,quasi}, we may define and study the functional $\mathcal E$ on the extended convex set
\begin{equation}
\S_\Lambda = \{Q\, | \, 0\leq Q + P^0 \leq 1, \, Q \in \gSt(\gH_\Lambda), \, \rho_Q \in {\mathcal{C}}
 \},
\end{equation}
where $\mathcal C$ is the so-called Coulomb space consisting of functions with finite Coulomb norm 
$$ \|\rho\|_{\mathcal{C}}^2:=D(\rho,\rho)=4\pi\int_{\R^3}\frac{|\widehat{\rho}(k)|^2}{|k|^2}dk.$$
More precisely, $\mathcal C$ is the Fourier inverse of the $L^2$ space with weight $1/|k|^2$. 

As our main result we obtain that, for any $\Lambda$,
$\E$ is bounded-below and has a minimizer on $\S_\Lambda $, therefore there exists a BDF-stable
vacuum.
\begin{theorem}\label{thm1}
Let $0\leq \alpha <4/\pi$, $n \in {\mathcal{C}}$. Then $\E$ satisfies, for any $Q\in\S_\Lambda$,
\begin{equation}
\E(Q)+\frac{\alpha}{2}D(n,n)\geq 0
\label{bounded_below}
\end{equation} 
and it is therefore bounded from below on $\S_\Lambda$. Moreover, there exists a minimizer $\bar{Q}$ of $\E$ on $\S_\Lambda$ such that 
$\bar{P}=\bar{Q} + P^0$ is a projector satisfying the
self-consistent equation
\begin{equation}
\label{sce}
\bar{P} = \chi_{(-\infty,0)}\Big(D^0 - \alpha \vp + \alpha\rho_{\bar{Q}} \ast \frac 1{|\cdot|} - \alpha
\frac{\bar{Q}(x,y)}{|\bx - \by|}\Big).
\end{equation}
Additionally, if $\alpha$ and $n$ satisfy
\begin{equation}
0\leq\alpha\frac\pi4\left\{1-\alpha\left(\frac\pi2 \sqrt{\frac{\alpha/2}{1-\alpha\pi/4}}+\pi^{1/6}2^{11/6}\right)\|n\|_{\mathcal{C}}\right\}^{-1}\leq1,
\label{condition_uniqueness}
\end{equation} 
then this global minimizer $\bar Q$ is unique and the associated polarized vacuum is neutral:
$$\langle\Omega_{\bar P}|\mathcal{Q}|\Omega_{\bar P}\rangle=\tr_{P^0}(\bar Q)=0.$$
\end{theorem}
The proof of this result is given in Section \ref{proof_section}.

Equation \eqref{sce} corresponds to Dirac's picture that
the ``correct'' vacuum $\bar P$ should be the projector
on the negative spectrum of an effective one-body
Hamiltonian. In the case without external
potential, $n = 0$, the free projector $P^0$ solves \eqref{sce} and is the unique BDF-stable vacuum \cite{CIL,BBHS}.

Numerically the self-consistent solution of \eqref{sce}
could be evaluated by a fixed point algorithm, starting with $P^0$.
In \cite{HLS1} we proved the convergence of this algorithm to a BDF-stable vacuum solving \eqref{sce},
under reasonable restrictions of the form $\alpha\sqrt{\|n\|_{L^2}^2+\|n\|^2_\mathcal{C}}\leq C_1$ and $\alpha \sqrt{\log\Lambda} \leq C_2$, using
the Banach fixed point theorem. This proof
is much more constructive than the direct variational proof which is given in Section \ref{proof_section}.
However, the result of \cite{HLS1} is local in the sense that it is valid for weak external potentials $\varphi=n\ast1/|\cdot|$ only. 

The condition \eqref{condition_uniqueness} means that if the overall charge of
the nuclei is not too big  and $\alpha$ is small enough, the BDF-stable vacuum is unique and 
stays neutral, cf. \cite{GMR,H}. In general, the solution found in Theorem \ref{thm1} can correspond to a charged vacuum.

There is an interesting symmetry property of the solutions of \eqref{sce} when $n$ is replaced by $-n$. Namely, if $P$ is a solution of \eqref{sce} with external density $n$, then $P'=Q'+P^0$ is a solution of \eqref{sce} with external density $-n$, where $Q'=-CQC^{-1}$, $C$ being the charge conjugation operator \cite[page 14]{Thaller}. The two dressed vacua $P$ and $P'$ have the same BDF energies and satisfy $\rho_{Q'}=-\rho_Q$, as suggested by the intuition. For this symmetry between matter and antimatter to be true, it is essential to have the Fermi level at $0$ and not at $-1$ (see, e.g., the comments of \cite[page 197]{SS} about this fact).

\bigskip

In Theorem \ref{thm1}, the cut-off $\Lambda$ can be chosen arbitrarily large and it is therefore natural to describe the behaviour of our solution as $\Lambda\to\infty$. 
\begin{theorem}\label{large_cutoff} 
Let be $n\in \mathcal{C}\cap L^2(\R^3)$ and $0\leq\alpha<4/\pi$. Then the solution $\bar Q_\Lambda=\bar P_\Lambda-P^0$ obtained in Theorem \ref{thm1} satisfies
$$\|{|D^0|^{1/2}\bar Q_\Lambda}\|_{\gS_2}\to 0,\qquad \alpha\|{\rho_{\bar Q_\Lambda}-n}\|_{\mathcal{C}}\to0$$
as $\Lambda\to\infty$, and therefore
\begin{equation}
\lim_{\Lambda\to\infty}\min_{\S_\Lambda}\E=-\frac{\alpha}{2}D(n,n).
\label{limit_min}
\end{equation} 
\end{theorem} 
 
In words, when $\Lambda\to\infty$, the vacuum polarization density \emph{totally cancels the external density $n$}, for $\rho_{\bar Q_\Lambda}\to n$ in $\mathcal{C}$. But since $\bar Q_\Lambda=P_\Lambda-P^0\to0$, this means that in the limit $\Lambda\to\infty$, $\bar Q_\Lambda$ and $\rho_{\bar Q_\Lambda}$ become independent. Therefore, the minimization without cut-off makes no sense both from a mathematical and physical point of view. Indeed \eqref{limit_min} easily implies that when no cut-off is imposed and when $\phi\neq0$, the infimum of the functional $\mathcal{E}$ is \emph{not attained}.
In physics, this ``nullification" of the theory as the cut-off $\Lambda$ diverges has been first suggested by Landau {\it et al.} \cite{Lan86,Lan84,Lan89,Lan100} and later studied by Pomeranchuk {\it et al.} \cite{Pom}.

In the next section, we propose a renormalization procedure in which we show an inequality of the form $\frac{2}{3\pi}\alpha_{\rm r}\log\Lambda\leq1$ where $\alpha_{\rm r}$ is the physical (renormalized) coupling constant, different from $\alpha$. With the usual value $\alpha_{\rm r}\simeq \frac{1}{137}$, this leads to the physical bound $\Lambda\leq 10^{280}$ (in units of $mc^2$).

The proof of Theorem \ref{large_cutoff} is given in Section \ref{proof_section}.

\medskip

\noindent {\bf Remark.}
If $n$ is smooth enough, it can be shown that
$$\|{|D^0|^{1/2}\bar Q_\Lambda}\|_{\gS_2}\leq C_1(\log\Lambda)^{-1},\qquad \alpha\|{\rho_{\bar Q_\Lambda}-n}\|_{\mathcal{C}}\leq C_2(\log\Lambda)^{-1}$$
for some constants $C_1$ and $C_2$.

\section{Reduced energy functional and charge renormalization}
Recall up to now the charge was kept to be the bare one.
Next we want to
derive a renormalization scheme consistent to any order of $\alpha$ for the solution of our minimization problem. Note that this procedure is well known in perturbation theory, see e.g. \cite[page 194]{RGA} and \cite{Hamm}.

We first simplify our BDF energy by neglecting the exchange term,
\begin{equation}
\label{reden}
\Er(Q)=\str_{P^0}(D^0 Q)-\alpha \int\rho_Q \vp +\frac{\alpha}{2}
D(\rho_Q,\rho_Q).
\end{equation}
From a physical point of view this is quite natural,
since the exchange term is usually treated together 
with a term describing the interaction with
the photon field to form the standard electron
{\em self-energy} that is a subject of the mass renormalization. 

Notice that since $\Er\geq \E$, the energy functional $\Er$ is obviously bounded from below on $\S_\Lambda$, by Theorem \ref{thm1}. We now state our
\begin{theorem}
Let $0\leq \alpha < 4/\pi$, $n \in {\mathcal{C}}$. Then $\Er$ possesses a minimizer $\bar{Q}$ on $\S_\Lambda$, which satisfies
\begin{equation}
\label{rsce0}
\bar Q = \chi_{(-\infty,0)}\Big(D^0 - \alpha \vp + \alpha\rho_{\bar Q} \ast \frac 1{|\cdot|} \Big) - P^0+\gamma_0,
\end{equation}
where $\gamma_0$ is a finite rank operator of the form
$$\gamma_0=\sum_{i=1}^Kn_i |\phi_i\rangle\langle\phi_i|,\qquad 0\leq n_i\leq1,$$
$(\phi_i)_{i=1}^K$ being an orthonormal basis of ${\rm ker}(D^0-\alpha\vp + \alpha\rho_{\bar Q} \ast 1/{|\cdot|})$.

\medskip

Additionally, if $\alpha$ and $n$ satisfy 
\begin{equation}
\alpha\pi^{1/6}2^{11/6}\|n\|_{\mathcal{C}}<1,
\label{cond_alpha_n_2}
\end{equation} then this global minimizer $\bar Q$ is unique and 
$${\rm ker}(D^0-\alpha\vp + \alpha\rho_{\bar Q} \ast 1/{|\cdot|})=\{0\}$$
which implies
\begin{equation}
\label{rsce}
\bar Q = \chi_{(-\infty,0)}\Big(D^0 - \alpha \vp + \alpha\rho_{\bar Q} \ast \frac 1{|\cdot|} \Big) - P^0.
\end{equation}
\end{theorem}

The proof is much simpler than the one of Theorem
\ref{thm1}: $\Er$ is now a coercive and convex continuous functional which is therefore weakly lower semi-continuous 
on the closed convex set $\S_\Lambda$, and possesses a minimizer. The proof that it satisfies the self-consistent equation \eqref{rsce0} is the same as the one of Theorem \ref{thm1}, except that due to the absence of the exchange term, one is not always able to prove that $\bar Q+P^0$ is a projector, as usual in reduced Hartree-Fock type theories \cite{Solovej}.

\bigskip

In order to perform
our renormalization scheme we expand \eqref{rsce} in powers
of $\alpha$. Assuming that \eqref{cond_alpha_n_2} holds, $0$ is not in the spectrum of the mean-field operator $D^\vp + \alpha\rho_{\bar Q} \ast 1/{|\cdot|}$ and we can use the resolvent representation \cite[Section VI, Lemma 5.6]{K} to
derive from \eqref{rsce} the self-consistent equation for
the VP-density $\rho_Q(\bx) = \Tr Q(\bx,\bx)$
\begin{equation}
\rho_Q(x) = -\frac 1{2\pi} \dK \, \Tr \left[ \frac 1{D^0 - \alpha \vp + \alpha\rho_Q \ast \frac 1{|\cdot|}
+ i\eta}
- \ide \right](x,x).
\end{equation}
Applying the resolvent equation $$ \frac 1{A-\alpha B} - \frac 1{A} = \alpha \frac 1{A} B \frac 1{A}
+ \alpha^2 \frac 1{A} B \frac 1{A} B \frac 1{A} + \alpha^3 \frac 1{A} B \frac 1{A} B \frac 1{A}
B \frac 1{A-\alpha B}$$
and using
Furry's Theorem \cite{Furry}, telling us that the corresponding $\alpha^2$-term with two potentials
vanish, we obtain
\begin{equation}\label{rscer}
\rho_Q = \alpha F_1[\rho_Q - n] +   F_{3}[\alpha\rho_Q - \alpha n]
\end{equation}
with 
\begin{multline*}
F_{3} [\rho](x) = \\ \dK\, \Tr \left[\ide \rho\ast \frac 1{|\cdot|} \ide \rho\ast \frac 1{|\cdot|}
\ide\rho\ast \frac 1{|\cdot|} \frac 1{D^0 - \alpha \rho + i\eta }\right](x,x).
\end{multline*} 

As realized first by Dirac \cite{D1,D2} and Heisenberg \cite{Hei}, cf. also \cite{FO},
the term $F_1[\rho]$ plays a particular role since it is
logarithmically ultraviolet divergent. Following, e.g., Pauli-Rose \cite{PauliRose},
one evaluates in Fourier representation $$\widehat F_1[\rho](k)=-\hat\rho(k)B_\Lambda(k),$$ with
\begin{equation}
B_\Lambda(k) =\frac1\pi \int_0^{\frac{\Lambda}{\sqrt{1+\Lambda^2}}}\frac{z^2-z^4/3}{1-z^2}\frac{dz}{1+|k|^2(1-z^2)/4},\label{pre}
\end{equation}
which can be decomposed into \cite[Equ. (5)-(9)]{PauliRose}
$ B_\Lambda(k)= B_\Lambda -  C_\Lambda(k)$, with
\begin{equation}
B_\Lambda = B_\Lambda(0) = \frac 1\pi \int_0^{\frac{\Lambda}{\sqrt{1+\Lambda^2}}}\frac{z^2-z^4/3}{1-z^2}\,dz
=\frac{2}{3\pi}\log(\Lambda)-\frac{5}{9\pi}+\frac{2}{3\pi}\log 2 + O(1/\Lambda^2) .
\end{equation}
and 
\begin{equation}
 \lim_{\Lambda \to \infty} C_{\Lambda}(k) = C(k)=-\frac 1{2\pi} \int_0^1 dx (1-x^2)\log[1+k^2(1-x^2)/4],
\end{equation}
which was first calculated by Serber and Uehling \cite{Se,Ue}. 

Denote $\rho= \rho_Q - n$ the total density, then
\eqref{rscer} reads in terms of $\rho$
\begin{equation}\label{r1}
\hat \rho + \hat n=  -\alpha B_\Lambda \hat \rho  + \alpha
C_\Lambda (k) \hr +  \widehat F_{3} [\alpha\rho],
\end{equation}
or equivalently
\begin{equation}\label{r1bis}
\alpha\hat \rho =-\alpha\hat n  -\alpha^2 B_\Lambda \hat \rho  + \alpha^2
C_\Lambda (k) \hr +  \alpha\widehat F_{3} [\alpha\rho]
\end{equation}
and
\begin{equation}\label{r2}
\alpha\hat \rho =-\frac \alpha{1+\alpha B_\Lambda } \hat n   + \frac \alpha{1+\alpha B_\Lambda}
C_\Lambda (k)\alpha \hr + \frac \alpha{1+\alpha B_\Lambda } \widehat F_{3} [\alpha\rho].
\end{equation}
To perform our renormalization scheme we fix as physical
(renormalized) objects $\alr\rr = \alpha \rho$,  with (cf. \cite[Equ. (7-18)]{IZ})
\begin{equation}
\alr = \frac \alpha{1+\alpha B_\Lambda}.
\label{ren_alpha}
\end{equation} 
Therefore we can
rewrite the self-consistent
equation \eqref{r1} as
\begin{equation}\label{r3}
\alr\hat \rho_{\mathrm{r}} =-\alr \hat n   +  \alpha_{\mathrm{r}}^2
C_\Lambda (k) \hat \rho_{\mathrm{r}} +  \alpha_{\mathrm{r}} \widehat F_{3} [\alr \rr],
\end{equation}
independently of  the bare $\alpha$. This equation uniquely defines
the VP density only depending on the physical
observable $\alr$, which is what we understand under consistent to any order. The
$\alr$ represents the dressed coupling constant, which
is observed in experiment and whose value is approximately $1/137$. Notice that
from formula \eqref{ren_alpha}, it follows that
necessarily $\alr B_\Lambda < 1$ and $\alr B_\Lambda \to 1$ as $\Lambda\to\infty$. We emphasize that although in the literature the expression of $\alr$ is sometimes expanded to get $\alr\simeq \alpha(1-\alpha B_\Lambda)$ leading to the condition $\alpha B_\Lambda<1$, the real constraint indeed applies to the physically observed $\alr$ and not the bare one.

Notice that equation \eqref{r3} satisfied by $\alpha_{\rm r}\rho_{\rm r}$ is exactly the same as equation \eqref{r1bis} satisfied by $\alpha\rho$, but with the logarithmically divergent term $\alpha^2 B_\Lambda \hat \rho$ dropped. Therefore, as usual in QED \cite{Dy}, the charge renormalization allows to simply justify the dropping of the divergent terms in the self-consistent equation. In practice \cite{MPS}, one would solve \eqref{r3} with $\alpha_{\rm r}\simeq 1/137$ and with $C_\Lambda(k)$ replaced by its limit $C(k)$.

\medskip

Returning to
the {\em effective} Hamiltonian $D^0 - \alpha\vp +\alpha \rho_Q\ast1/|\cdot|$ and
inserting \eqref{r3}, i.e.
expressing in terms of the physical
objects, we obtain
\begin{equation}\label{effd}
D^0 + \alr \rr\ast \frac 1{|\cdot|} = D^0 - \alr n\ast\frac 1{|\cdot|}
+V_{{\rm eff}},
\end{equation}
with $$V_{{\rm eff}} =\frac 2{\pi^3} \F^{-1}\left[\frac {\alpha_{\mathrm{r}}^2 C_\Lambda (k)\hat 
\rho_{\mathrm{r}} (k)
+ \alr \hat F_3(\alr \rr)}{ k^2}\right](x)$$
the effective self-consistent potential, where $\F^{-1}$ denotes the inverse Fourier transform.
Notice, this equation is valid for any strength of the external potential.
However, expanding $\rr$ in $\alr$, 
we obtain to lowest order in $\alr$
\begin{eqnarray*}
 V_{{\rm eff}} &\simeq&
\alpha_{\mathrm{r}}^2 \frac 2{\pi^3} \F^{-1}\left[\frac {C_\Lambda (k)\hat n (k)}{ k^2}\right](x) \\
&\simeq&
\frac {\alpha_{\mathrm{r}}^2 }{3\pi} \int_1^\infty dt (t^2-1)^{1/2}\left[\frac 2{t^2} + \frac 1{t^4}\right] \int dx'
e^{-2|x-x'|t} \frac {n(x')}{|x-x'|},
\end{eqnarray*}
the Uehling potential \cite{BR}. Concerning a point like particle this
potential was first written down in a closed form by Schwinger \cite{Sch}.
The next term in $V_{{\rm eff}}$ is of order $\alr (\alr Z)^3$.
In principle all higher order corrections
can be evaluated explicitly, which is not the task of our paper.

Finally we note that the convergence of the
term in the right hand side of \eqref{r1}, in the case of the VP-density in the Furry picture,
i.e. $\alpha \widehat F_{3} (\alpha n)$, was shown in various papers. The most clarifying proof with respect to
spurious third order contributions can probably be found
in \cite{SM} (for earlier proofs, in particular corresponding
to muonic atoms, we refer to the references in \cite{SM}). However the fact that this term,
$\alpha \widehat F_{3} (\alpha n)$, additionally gives rise
to a well defined self-adjoint operator was recently proved in \cite{HS}.

\section{Proof of Theorems \ref{thm1} and \ref{large_cutoff}}\label{proof_section}
In this last Section, we give the proof of our main Theorems. 

\subsection{Proof of Theorem \ref{thm1}}
The proof that $\E$ is well-defined on $\S_\Lambda$ can be found in details in \cite[Theorem 1]{HLS1}. For simplicity, we extend $\E$ to the closed convex set 
$$\S'_\Lambda=\{Q\in\gS_2(\gH_\Lambda),\ 0\leq Q+P^0\leq 1,\ \rho_Q\in\mathcal{C}\}$$
of the Hilbert space $\mathcal{H}:=\{Q\in\gS_2(\gH_\Lambda),\ \rho_Q\in\mathcal{C}\}$, by simply letting
$$\E(Q)=F(Q)-\alpha D(\rho_Q,n) +\frac{\alpha}{2} D(\rho_Q,\rho_Q),$$
\begin{equation}
F(Q):= \left\{\begin{array}{ll}
\str_{P^0}(D^0Q)-\frac{\alpha}{2}\iint\frac{|Q(x,y)|^2}{|\bx-\by|}dx\,dy & {\rm if\ } Q\in\gS_1^{P^0}(\gH_\Lambda)\\
+\infty & {\rm if\ } Q\notin\gS_1^{P^0}(\gH_\Lambda)
\end{array} \right.
\label{function_F}.
\end{equation} 
Let us recall the inequality established in \cite{BBHS}
$$F(Q)\geq (1-\alpha\pi/4)\str_{P^0}(D^0Q)=(1-\alpha\pi/4)(\tr(|D^0|Q^{++})-\tr(|D^0|Q^{--}))$$
(notice that $Q^{++}\geq0$ and $Q^{--}\leq0$ when $Q\in\S_\Lambda'$), which easily implies the bound \eqref{bounded_below} since 
\begin{equation}
\label{bound_below_E}
\E(Q)\geq (1-\alpha\pi/4)\str_{P^0}(D^0Q) +\frac{\alpha}{2}\|\rho_Q-n\|_\mathcal{C}^2-\frac{\alpha}{2}\|n\|_\mathcal{C}^2\geq -\frac{\alpha}{2}\|n\|_\mathcal{C}^2.
\end{equation} 
This also easily shows that both $F$ and $\E$ are strongly lower semi-continuous and coercive on $\S'_\Lambda$. We now prove that $\E$ is indeed weakly lower semi-continuous (wlsc) on $\S_\Lambda'$ in $\mathcal{H}$, which will show the existence of a minimizer since $\S'_\Lambda$ is closed and convex, and therefore weakly closed.

\medskip

{\it Step 1: $\E$ is wlsc on $\S'_\Lambda$}. Since the functional
$$Q\mapsto-\alpha D(\rho_Q,n) +\frac{\alpha}{2} D(\rho_Q,\rho_Q)=\frac{\alpha}{2}\|\rho_Q-n\|_\mathcal{C}^2-\frac{\alpha}{2}\|n\|_\mathcal{C}^2$$
is easily seen to be wlsc on the convex set $S'_ \Lambda$, it only remains to prove that $F$ (defined in \eqref{function_F}) is wlsc on $S'_ \Lambda$. To this end, we consider a weakly converging sequence $Q_n\wto Q$ in $\mathcal{H}$, such that $Q_n\in\S'_\Lambda$ for each $n$. If $\liminf_nF(Q_n)=\infty$, there is nothing to show and we can therefore assume that $(Q_n^{++})_{n\geq1}$ and $(Q_n^{--})_{n\geq1}$ are bounded in $\gS_1(\gH_\Lambda)$. Due to the cut-off $\Lambda$ in Fourier space, $(Q_n(x,y))_{n\geq1}$ is bounded in the Sobolev space $H^1(\R^6,\cz^4\otimes\cz^4)$, $(\rho_{|D^0|Q^{++}_n})_{n\geq1}$ and $(\rho_{|D^0|Q^{--}_n})_{n\geq1}$ are bounded for instance in $H^1(\R^3,\R)$. We may thus assume, up to a subsequence, that $Q_n(x,y)\to Q(x,y)$ in $L^2_{\rm loc}(\R^6,\cz^4\otimes\cz^4)$, that $\rho_{|D^0|Q^{++}_n}\to\rho_{|D^0|Q^{++}}$ and $\rho_{|D^0|Q^{--}_n}\to\rho_{|D^0|Q^{--}}$ in $L^1_{\rm loc}(\R^3,\R)$.

Let us now consider two real functions $\eta,\xi\in\mathcal{C}^\infty([0;\infty);[0;1])$ such that $\eta(t)=1$ if $t\in[0;1]$, $\eta(t)=0$ if $t\geq2$, $0\leq \eta(t)\leq1$ if $t\in[1;2]$, and $\eta^2+\xi^2=1$. We now define $\eta_R(\bx):=\eta(|\bx|/R)$ and $\xi_R(\bx):=\xi(|\bx|/R)$ for $\bx\in\R^3$. In the following, we also denote by $\eta_R$ and $\xi_R$ the multiplication operators by the functions $\eta_R$ and $\xi_R$, acting on $\gH_\Lambda$.
\begin{lemma}\label{estim_decay_D0} We have $\left\|\, [\xi_R,|D^0|]\,\right\|_{\gS_\infty(\gH_\Lambda)}=O(1/R)$.
\end{lemma} 
\begin{proof}
We compute $\langle\psi|[\xi_R,|D^0|]|\chi\rangle$ in Fourier space, for some $\psi,\chi\in\gH_\Lambda$ (we use the notation $E(p)=\sqrt{1+p^2}$):
\begin{eqnarray*}
\langle\psi|[\xi_R,|D^0|]|\chi\rangle & = & \iint_{\R^6}\widehat{\xi_R}(p-q)\overline{\widehat{\psi}(p)}\widehat{\chi}(q)\left(E(q)-E(p)\right)dp\,dq\\
 & = & \iint_{\R^6}\widehat{\xi_R}(r)\overline{\widehat{\psi}\left(s+\frac{r}{2}\right)}\widehat{\chi}\left(s-\frac{r}{2}\right)\left(E\left(s-\frac{r}{2}\right)-E\left(s+\frac{r}{2}\right)\right)ds\,dr
\end{eqnarray*} 
and therefore, using the inequality $|E(x)- E(x-y)|\leq|y|$, we obtain
$$ \left|\langle\psi|[\xi_R,|D^0|]|\chi\rangle\right| \leq \left(\int_{\R^3}\left|r\widehat{\xi_R}(r)\right|dr\right)\|\widehat{\psi}\|_{L^2} \|\widehat{\chi}\|_{L^2}$$
and
$$\left\|\, [\xi_R,|D^0|]\,\right\|_{\gS_\infty(\gH_\Lambda)}\leq \int_{\R^3}\left|r\widehat{\xi_R}(r)\right|dr=\frac{C}{R}\int_{\R^3}\left|r\widehat{\xi_1}(r)\right|dr.$$
\end{proof} 
Using this Lemma, we may now write
\begin{eqnarray*}
\tr(|D^0|Q_n^{++}) & =& \tr(\eta_R^2|D^0|Q_n^{++})+\tr(\xi_R^2|D^0|Q_n^{++})\\
 &=& \tr(\eta_R|D^0|Q_n^{++}\eta_R)+\tr(|D^0|\xi_RQ_n^{++}\xi_R)+\tr([\xi_R,|D^0|]Q_n^{++}\xi_R)\\
 & = & \tr(\eta_R|D^0|Q_n^{++}\eta_R)+\tr(|D^0|\xi_RQ_n^{++}\xi_R)+O(1/R)
\end{eqnarray*} 
since 
$$|\tr([\xi_R,|D^0|]Q_n^{++}\xi_R)|\leq \left\| [\xi_R,|D^0|]\right\|_{\gS_\infty(\gH_\Lambda)}\left\|Q_n^{++}\right\|_{\gS_1(\gH_\Lambda)}=O(1/R)$$
by Lemma \ref{estim_decay_D0} and since by assumption $(Q_n^{++})_{n\geq1}$ is bounded in $\gS_1(\gH_\Lambda)$. With the same argument for $Q_n^{--}$, we obtain
\begin{multline*}
\str_{P^0}(D^0Q_n)=\tr(\eta_R|D^0|Q_n^{++}\eta_R)-\tr(\eta_R|D^0|Q_n^{--}\eta_R)\\
+\tr(|D^0|\xi_RQ_n^{++}\xi_R)-\tr(|D^0|\xi_RQ_n^{--}\xi_R)+O(1/R).
\end{multline*} 
On the other hand, we have 
\begin{multline*}
\iint\frac{|Q_n(x,y)|^2}{|\bx-\by|}dx\,dy=\iint\frac{\eta_R(\bx)^2\eta_{3R}(\by)^2|Q_n(x,y)|^2}{|\bx-\by|}dx\,dy\\+\iint\frac{\xi_R(\bx)^2|Q_n(x,y)|^2}{|\bx-\by|}dx\,dy+O(1/R)
\end{multline*} 
since 
$$\iint\frac{\eta_R(\bx)^2\xi_{3R}(\by)^2|Q_n(x,y)|^2}{|\bx-\by|}dx\,dy\leq \frac{\|Q_n\|^2_{\gS_2(\gH_\Lambda)}}{R}.$$
We therefore obtain
\begin{multline*}
F(Q_n)=\tr(\eta_R|D^0|Q_n^{++}\eta_R)-\tr(\eta_R|D^0|Q_n^{--}\eta_R)\\-\frac{\alpha}{2}\iint\frac{\eta_R(\bx)^2\eta_{3R}(\by)^2|Q_n(x,y)|^2}{|\bx-\by|}dx\,dy
+\tr(|D^0|\xi_RQ_n^{++}\xi_R)-\tr(|D^0|\xi_RQ_n^{--}\xi_R)\\-\frac{\alpha}{2}\iint\frac{\xi_R(\bx)^2|Q_n(x,y)|^2}{|\bx-\by|}dx\,dy+O(1/R).
\end{multline*} 
Notice now that $0\leq Q_n+P^0\leq 1$ implies $|Q_n|^2\leq Q_n^{++}-Q_n^{--}$ (see \cite{BBHS}). We now localize this inequality to obtain $\xi_R|Q_n|^2\xi_R\leq \xi_RQ_n^{++}\xi_R-\xi_RQ_n^{--}\xi_R$. By Kato's inequality \cite{BBHS}, we now have
\begin{eqnarray*}
\iint\frac{\xi_R(\bx)^2|Q_n(x,y)|^2}{|\bx-\by|}dx\,dy & \leq & \frac{\pi}{2}\tr(|D^0|\xi_RQ_n^2\xi_R)\\
 & \leq & \frac{\pi}{2}(\tr(|D^0|\xi_RQ_n^{++}\xi_R)-\tr(|D^0|\xi_RQ_n^{--}\xi_R))
\end{eqnarray*} 
and therefore, since $(1-\alpha\pi/4)\geq0$, 
\begin{multline*}
F(Q_n)\geq\tr(\eta_R|D^0|Q_n^{++}\eta_R)-\tr(\eta_R|D^0|Q_n^{--}\eta_R)\\-\frac{\alpha}{2}\iint\frac{\eta_R(\bx)^2\eta_{3R}(\by)^2|Q_n(x,y)|^2}{|\bx-\by|}dx\,dy+O(1/R).
\end{multline*}
Passing now to the limit as $n\to\infty$ and using the local compactness of $Q_n(x,y)$ in $L^2_{\rm loc}(\R^6)$ and $\rho_{|D^0|Q_n^{++}}$, $\rho_{|D^0|Q_n^{--}}$ in $L^1_{\rm loc}(\R^3)$, we obtain
\begin{multline*}
\liminf_{n\to\infty}F(Q_n)\geq\tr(\eta_R|D^0|Q^{++}\eta_R)-\tr(\eta_R|D^0|Q^{--}\eta_R)\\-\frac{\alpha}{2}\iint\frac{\eta_R(\bx)^2\eta_{3R}(\by)^2|Q(x,y)|^2}{|\bx-\by|}dx\,dy+O(1/R).
\end{multline*}
If we now let $R\to\infty$, we obtain $\liminf_{n\to\infty}F(Q_n)\geq F(Q)$ and therefore $F$ is wlsc on $\S_\Lambda'$.

\medskip

{\it Step 2: at least one of the minimizers satisfies \eqref{sce}.} In the previous step, we have shown the existence of a minimizer. It now remains to show that one of them indeed satisfies \eqref{sce}.
\begin{lemma}
\label{eq_minimizer}
Let $\bar Q$ be a minimizer of $\E$ in $\S_\Lambda'$. Then either $\bar Q+P^0$ is a projector, or
\begin{equation}
\bar Q+P^0=\bar P+\mu|f\rangle\langle f|,
\label{form_lemma}
\end{equation} 
where $\bar P$ is a projector, $\mu\in(0;1)$ and $f\in\ker(D_{\bar Q})$, with
$$D_{\bar Q}:=D^0 - \alpha \vp + \alpha\rho_{\bar Q} \ast \frac 1{|\cdot|} - \alpha\frac{\bar Q(x,y)}{|\bx - \by|}.$$
\end{lemma} 
\begin{proof}
Our proof is inspired by classical arguments already used in the Hartree-Fock theory \cite{Lieb,Bach}.

Notice that since $\bar Q$ is compact, $\bar Q+P^0$ is a compact perturbation of $P^0$ and therefore its essential spectrum is $\sigma_{\rm ess}(\bar Q+P^0)=\{0,1\}$, meaning that $\sigma(\bar Q+P^0)\cap(0;1)$ only contains eigenvalues of finite multiplicity accumulating at $\{0,1\}$. Let us assume that $\bar Q+P^0$ possesses two different eigenvectors $\phi_1$, $\phi_2$:
$$\bar Q+P^0=\lambda_1|\phi_1\rangle\langle\phi_1|+\lambda_2|\phi_2\rangle\langle\phi_2|+G$$
where $\lambda_1,\lambda_2\in(0;1)$ and $G\phi_1=G\phi_2=0$. We now introduce $\bar Q_\epsilon:=\bar Q+\epsilon|\phi_1\rangle\langle\phi_1|-\epsilon|\phi_2\rangle\langle\phi_2|$ which belongs to $\S_\Lambda$ for $\epsilon$ small enough and compute
$$\E(\bar Q_\epsilon)=\E(\bar Q)+\epsilon\big(\langle\phi_1|D_{\bar Q}|\phi_1\rangle-\langle\phi_2|D_{\bar Q}|\phi_2\rangle\big)-\epsilon^2\frac{\alpha}{2}\iint\frac{|\phi_1\wedge\phi_2(x,y)|^2}{|\bx-\by|}dx\,dy.$$
Therefore, using either the first order term in $\epsilon$ if it does not vanish, or the second order term, we can always decrease the energy. This is a contradiction which implies that $\sigma(\bar Q+P^0)\cap(0;1)$ contains at most one eigenvalue of multiplicity 1 and thus
$$\bar Q+P^0=\bar P+\mu|f\rangle\langle f|$$
where $\bar P$ is a projector and $\mu\in[0;1)$. If $\mu\neq0$, using the same type of variation $\bar Q_\epsilon:=\bar Q+\epsilon|f\rangle\langle f|$, we easily show that indeed $f\in\ker(D_{\bar Q})$. 
\end{proof} 

If $\bar Q$ is a minimizer of the form \eqref{form_lemma}, we now see that 
\begin{eqnarray*}
\E(\bar Q) & = & \E(\bar P-P^0)+\mu\langle D_{\bar P-P^0}f,f\rangle\\
 &= &\E(\bar P-P^0)+\mu\langle D_{\bar Q}f,f\rangle\\
 & = & \E(\bar P-P^0)
\end{eqnarray*}  
and therefore $\bar P-P^0$ is also a minimizer of $\E$ (i.e. $\bar P$ is BDF-stable vacuum). In \cite[proof of Theorem 2]{HLS1}, we have already shown that a minimizer of $\E$ on $\S_\Lambda$ taking the form $\bar P-P^0$ where $\bar P$ is an orthogonal  projector, is indeed a solution of the self-consistent equation \eqref{sce}.

\medskip

{\it Step 3: uniqueness of the global minimizer of $\E$ under the condition \eqref{condition_uniqueness}.} Due to \cite[Theorem 2]{HLS1}, we know that the global minimizer $\bar Q$ of $\E$ is unique if $D_{\bar Q}$ satisfies
$$d|D_{\bar Q}|\geq|D^0|$$
for some $d$ such that $\alpha d\pi/4\leq1$.

We know that $\E(\bar Q)\leq \E(0)=0$ and therefore, by an argument similar to \eqref{bound_below_E},
\begin{equation}
\left(\frac2\pi-\frac\alpha2\right)\iint_{\R^6}\frac{|\bar Q(x,y)|^2}{|x-y|}dx\,dy +\frac\alpha2\|\rho_{\bar Q}-n\|_{\mathcal{C}}^2\leq \frac\alpha2\|n\|^2_{\mathcal{C}}
\label{unif_estim_Q}
\end{equation} 
and thus
\begin{equation}
\iint_{\R^6}\frac{|\bar Q(x,y)|^2}{|x-y|}dx\,dy\leq \frac{\alpha\pi/4}{1-\alpha\pi/4}\|n\|_{\mathcal{C}}^2
\label{estim_Q1}
\end{equation} 
\begin{equation}
\|\rho_{\bar Q}-n\|_{\mathcal{C}}\leq \|n\|_{\mathcal{C}}.
\label{estim_Q2}
\end{equation} 
Recall that $D_{\bar Q}=D^0+\alpha\varphi'_{\bar Q}-\alpha R_{\bar Q}$ where $\varphi'_{\bar Q}=(\rho_{\bar Q}-n)\ast1/|\cdot|$ and $R_{\bar Q}$ is the operator with kernel $\bar Q(x,y)/|x-y|$. Now, we have
$$\left\|\phi'_{\bar Q}\frac{1}{|D^0|}\right\|_{\gS_\infty(\gH_\Lambda)} \leq \left\|\phi'_{\bar Q}\frac{1}{|D^0|}\right\|_{\gS_6(\gH_\Lambda)}\leq (2\pi)^{-1/2}\left\|\phi'_{\bar Q}\right\|_{L^6}\left\|E(\cdot)^{-1}\right\|_{L^6}$$
where we recall that $E(p)=\sqrt{1+p^2}$, and by \cite[Theorem 4.1]{Simon}. Therefore
$$\left\|\phi'_{\bar Q}\frac{1}{|D^0|}\right\|_{\gS_\infty(\gH_\Lambda)} \leq S_6\left\|\phi'_{\bar Q}\right\|_{L^6}\leq S_6C_6\left\|\nabla\phi'_{\bar Q}\right\|_{L^2}=(4\pi)S_6C_6\|\rho_{\bar Q}-n\|_{\mathcal{C}} $$
with $S_6=2^{-5/6}3^{1/6}\pi^{-1/6}$ and where $C_6=3^{-1/6}2^{2/3}\pi^{-2/3}$ is the Sobolev constant for the inequality $\|f\|_{L^6(\R^3)}\leq C_6\|\nabla f\|_{L^2(\R^3)}$. Due to \eqref{estim_Q2}, this shows that
$$|\phi'_{\bar Q}|\leq \kappa\|n\|_{\mathcal{C}}\;|D^0|,$$
where $\kappa=(4\pi)S_6C_6=\pi^{1/6}2^{11/6}$.
On the other hand, we know from \cite[Proof of Lemma 4]{HLS1} that
$$|R_{\bar Q}|\leq \sqrt{\frac\pi2 \iint_{\R^6}\frac{|\bar Q(x,y)|^2}{|x-y|}dx\,dy}\; |D^0|$$
and therefore, using \eqref{estim_Q1},
$$|R_{\bar Q}|\leq \frac\pi2 \sqrt{\frac{\alpha/2}{1-\alpha\pi/4}}\|n\|_{\mathcal{C}}\; |D^0|.$$

As a conclusion, when 
$$\alpha\left(\frac\pi2 \sqrt{\frac{\alpha/2}{1-\alpha\pi/4}}+\kappa\right)\|n\|_{\mathcal{C}}<1,$$
$D_{\bar Q}$ fulfills $d|D_{\bar Q}|\geq|D^0|$ with
$$d=\left\{1-\alpha\left(\frac\pi2 \sqrt{\frac{\alpha/2}{1-\alpha\pi/4}}+\kappa\right)\|n\|_{\mathcal{C}}\right\}^{-1}.$$
Applying now \cite[Theorem 2]{HLS1}, we obtain that the minimizer $\bar Q$ is unique when $\alpha d \pi/4\leq 1$, i.e. under the condition \eqref{condition_uniqueness}.

Assuming now that \eqref{condition_uniqueness} holds, let us show that the unique BDF-stable vacuum $\bar P$ is not charged. To this end, we define, for $t\in[0;1]$,
$$\bar Q(t)=\chi_{(-\infty;0)}\left(D^0+\alpha t(\rho_{\bar Q}-n)\ast\frac{1}{|\cdot|}-\alpha t \frac{\bar Q(x,y)}{|x-y|}\right)-P^0.$$
$t\mapsto\bar Q(t)$ is a continuous function for the $\gS_2(\gH_\Lambda)$ topology, since by the previous estimates $D_{\bar Q}(t)=D^0+\alpha t(\rho_{\bar Q}-n)\ast\frac{1}{|\cdot|}-\alpha t \frac{\bar Q(x,y)}{|x-y|}$ possesses a gap around $0$, uniformly in $t\in[0;1]$. This implies that 
$$q:t\mapsto \tr_{P^0}(Q(t))=\tr(Q(t)^3)$$
is continuous on $[0;1]$, by \cite[Lemma 2]{HLS1}. Since $q(0)=0$ and $q(t)$ is an integer for any $t\in[0;1]$, we therefore deduce that
$$q(1)=\langle\Omega_{\bar P}|\mathcal{Q}|\Omega_{\bar P}\rangle=\tr_{P^0}(\bar Q)=0.$$
This ends the proof of Theorem \ref{thm1}.\qed

\subsection{Proof of Theorem \ref{large_cutoff}} We first prove \eqref{limit_min} which will easily imply that $\bar Q_\Lambda$ obtained by Theorem \ref{thm1} behaves at stated as $\Lambda\to\infty$, due to \eqref{bound_below_E}. To this end, we introduce
$$Q_\Lambda:=\chi_{(-\infty;0)}\left(D^0-\alpha n_\Lambda\ast\frac{1}{|\cdot|}\right)-P^0\in\S_\Lambda,$$
$$\widehat{n_\Lambda}(k):=\frac{\widehat{n}(k)}{1+\alpha B_\Lambda(k)}$$
where we recall that $B_\Lambda(k)=B_\Lambda-C_\Lambda(k)$  is defined in \eqref{pre}. We now show that
$$\lim_{\Lambda\to\infty}\E(Q_\Lambda)=-\frac\alpha2 D(n,n),$$
which will imply \eqref{limit_min}, by \eqref{bounded_below}.

Let us now compute $\rho_\Lambda:=\rho_{Q_\Lambda}$. By \eqref{r1}, $\rho_\Lambda$ satisfies 
$$\widehat{\rho_\Lambda}(k)=\frac{\alpha}{1+\alpha B_\Lambda(k)}\widehat{n}(k)B_\Lambda(k)+\widehat{F_3}[\alpha n_\Lambda](k),$$
and therefore
$$\widehat{\rho_\Lambda}(k)-\widehat{n}(k)=-\widehat{n_\Lambda}(k)+\widehat{F_3}[\alpha n_\Lambda](k).$$
When $\alpha>0$, since $(1+\alpha B_\Lambda(k))^{-1}\to0$ a.e., we obtain by Lebesgue's dominated convergence Theorem that $\|n_\Lambda\|_{\mathcal{C}\cap L^2}\to0$ as $\Lambda\to\infty$. By the fixed-point estimates of \cite{HLS1} in the case of the reduced model (they are then independent on the cut-off $\Lambda$ as this can be seen from the proof of \cite[Theorem 3]{HLS1}), it is known that $F_3$ is continuous at $0$ for the $\mathcal{C}\cap L^2$ topology. We therefore obtain
$$\lim_{\Lambda\to\infty}\|\rho_\Lambda-n\|_{\mathcal{C}\cap L^2}=0.$$
On the other hand, we also know from the bounds proved in \cite{HLS1}, that 
$$\str_{P^0}(D^0Q_\Lambda)^{1/2}=\tr(|D^0|Q^2_\Lambda)^{1/2}\leq C\alpha\|\rho_\Lambda-n\|_{\mathcal{C}\cap L^2}$$
for some constant $C$ independent of $\Lambda$. Therefore
$$\lim_{\Lambda\to\infty}\E(Q_\Lambda)=-\frac\alpha2 D(n,n)$$
which ends the proof of Theorem \ref{large_cutoff}.\qed

\bibliographystyle{amsplain}

\end{document}